\documentclass{PoS}

\title{Spectral Projectors Method for Staggered Fermions}

\ShortTitle{Spectral Projectors with Staggered Fermions}

\author{
	\speaker{Claudio Bonanno}\thanks{Corresponding author.} $^{a}$, Giuseppe Clemente$^{a}$, Massimo D'Elia$^{a}$ and Francesco Sanfilippo$^{b}$\\ \\
	\llap{$^a$}Universit\`a di Pisa and INFN Sezione di Pisa, Largo Pontecorvo 3, I-56127 Pisa, Italy\\
	\llap{$^b$}INFN Sezione di Roma 3, Via della Vasca Navale 84, I-00146 Roma, Italy\\ \\	
	E-mail: \email{claudio.bonanno@pi.infn.it}, \email{giuseppe.clemente@pi.infn.it}, \email{massimo.delia@unipi.it}, \email{sanfilippo@roma3.infn.it}
}

\abstract{We extend the spectral projectors method to staggered fermions. Applying the index theorem to the staggered Dirac operator it is possible to work out an expression for the topological susceptibility which depends only on the orthogonal projectors on quasi zero-modes, as it has already been done
for Dirac-Wilson fermions. Besides, we generalize this method deriving analogous expressions for all higher-order coefficients in the $\theta$-expansion of the vacuum energy.}

\FullConference{37th International Symposium on Lattice Field Theory - Lattice2019\\
		16-22 June 2019\\
		Wuhan, China}

\usepackage{amsmath}
\usepackage{braket}
\usepackage{slashed}

\newcommand{\beq}{\begin{eqnarray}}
\newcommand{\eeq}{\end{eqnarray}}

\DeclareMathOperator{\Tr}{Tr}
\DeclareMathOperator{\Index}{Index}

\graphicspath{{./figures/}}

\begin{document}

\section{Introduction}
\label{intro}

The topological properties of the gauge sector of QCD and QCD-like theories, and the related $\theta$-dependence, is one of the most extensively studied non-perturbative features of these models and Monte Carlo simulations on the lattice are the most natural choice to investigate them.

The continuum definition of the topological charge in Yang-Mills theories can be expressed in terms of gluon fields as
\beq\label{continuum_charge}
Q = \int d^4x \frac{1}{64 \pi^2} \epsilon_{\mu\nu\rho\sigma}F^a_{\mu\nu}(x)F^a_{\rho\sigma}(x),
\eeq
and is integer valued when proper boundary conditions are chosen (such as periodic ones on a finite volume). This quantity can also be related to fermion field properties via the index theorem:
\beq\label{indexth}
Q = \Index\{\slashed{D}\} = n_+-n_-=\Tr\{\gamma_5\},
\eeq
where $n_{\pm}$ are, respectively, the number of left-handed and right-handed zero-modes of $\slashed{D}$. On the lattice, many definitions of topological charge can be assigned, all agreeing in the continuum limit but with different discretization corrections.

Gluonic definitions are defined rewriting Eq.~(\ref{continuum_charge}) in terms of lattice gauge links. Despite having the correct continuum limit, they are non-integer and their correlation functions are subject to multiplicative and additive renormalizations, which have to be properly treated (e.g.~by using smoothing methods such as cooling or gradient flow to damp UV fluctuations).

Fermionic definitions are instead based on a counting of zero-modes, cf.~Eq.~(\ref{indexth}), and in principle they can be obtained from the evaluation of $\Tr\{\gamma_5\}$ on a diagonal basis of the lattice Dirac operator. This however presents some difficulties related to the lattice implementation of chiral fermions. Indeed, while some particular discretizations (such as overlap fermions) satisfy an approximate chiral symmetry and possess exact chiral modes, some others (e.g. Wilson or staggered fermions) do not. Because of this, defining the lattice topological charge as the trace of the discretized $\gamma_5$ requires proper renormalizations~\cite{Bochicchio:1985xa, smit_and_vink, vink}. The spectral projectors method, which relies on the latter strategy, has been used to obtain a theoretically well-posed definition of the continuum topological susceptibility $\chi=\braket{Q^2}/V$ \cite{luscher_1,luscher_2} which is also easily adaptable for numerical simulations on the lattice. Up to now, this method has been defined for Dirac-Wilson fermions and successfully tested in pure Yang-Mills theories~\cite{luscher_3,Cichy_wilson_spectral_quenched} and in full QCD~\cite{athenodorou_wilson_spectral_fullQCD}.

Since all the methods exposed above yield the same results when the continuum limit is taken, which method should be adopted can be decided considering two main issues: computational cost and the magnitude of corrections to the continuum limit. The second aspect can be particularly relevant when dealing with dynamical fermions with light masses. In the continuum, zero-modes of $\slashed{D}$ suppress configurations with non-zero charge; on the lattice, instead, the presence of large would-be-zero-modes fails to efficiently suppress such configurations, resulting in larger values of $\chi$. This makes the extrapolation towards the continuum particularly difficult, both at zero and finite temperature~\cite{Bonati:2015vqz,Petreczky:2016vrs,Borsanyi:2016ksw,Burger:2018fvb,Bonati:2018blm}.

A recently-proposed heuristic solution is to reweight configurations by hand using the continuum lowest eigenvalues~\cite{Borsanyi:2016ksw}. Instead, a theoretically better founded solution could be to use a fermionic definition of the topological charge matching the same discretization of the sea quarks. This possibility is supported by a recent study~\cite{athenodorou_wilson_spectral_fullQCD}, where the topological susceptibility of QCD with twisted mass Wilson fermions is measured through twisted mass spectral projectors, resulting in an improved scaling towards the continuum compared to the standard gluonic measure.

In this paper we summarize the main ideas and results of \cite{Bonanno:2019xhg} (to which we refer for further details), where the spectral projectors definition of $\chi$ is extended to the case of staggered fermions, the main motivation being to 
adopt it in ongoing lattice investigations of $\theta$-dependence in full QCD with staggered fermions~\cite{Bonati:2018blm}. Besides, the method is also generalized to any higher-order cumulant of the charge distribution, which are related to higher-order coefficients of the $\theta$-expansion of the free energy. Given these aims, the shown numerical results will be limited to the quenched case, while results for full QCD will be presented in a separate work.

This paper is organized as follows: in Section~\ref{setup} we define spectral projectors in the case of staggered fermions, deriving spectral expressions for the topological susceptibility and for all higher-order cumulants, in Section~\ref{results} we present numerical results for the pure $SU(3)$ gauge theory and in Section~\ref{conclusions} we discuss conclusions and future perspectives.

\section{Spectral projectors method for staggered fermions}\label{setup}

\subsection{Spectral definition of topological observables}\label{subsec_stag_sp_chi}

The index theorem for the staggered Dirac operator $D_{\text{\textit{st}}}$ in $d$ space-time dimensions can be written considering that, in the continuum, it describes $2^{d/2}$ degenerate flavors: $Q_{0\text{\textit{st}}} = (-2)^{-d/2} \Tr\{\Gamma_5\}$, where $\Gamma_5$ is the staggered discretization of $\gamma_5$. The renormalization of the bare charge has been discussed in Refs.~\cite{smit_and_vink,vink}, where we refer the reader for further details. Making use of the anomalous Ward identities for staggered fermions, the renormalized staggered charge is
\beq
Q_{\text{\textit{st}}} = \frac{Z_P^{(\text{\textit{s}})}}{Z_S^{(\text{\textit{s}})}} Q_{0\text{\textit{st}}},
\label{q_stag_ren}
\eeq
where $Z_S^{(\text{\textit{s}})}$ and $Z_P^{(\text{\textit{s}})}$ refer, respectively, to the scalar and pseudo-scalar flavor-singlet bare densities $S_0=\bar{\psi}\psi$ and $P_0=\bar{\psi}\Gamma_5\psi$. Following the same line of reasoning of Ref.~\cite{luscher_2}, the ratio $Z_S^{(\text{\textit{s}})}/Z_P^{(\text{\textit{s}})}$ can be expressed as
\beq\label{stag_sp_Z_S_Z_P}
\left(\frac{Z_P^{(\text{\textit{s}})}}{Z_S^{(\text{\textit{s}})}}\right)^2 = \frac{\braket{\Tr\{\mathbb{P}_M\}}}{\braket{\Tr{\{\Gamma_5 \mathbb{P}_M\Gamma_5 \mathbb{P}_M\}}}},
\eeq
where $\mathbb{P}_M$ is the spectral projector on eigenspaces of $D_{\text{\textit{st}}}$ with eigenvalues $-i \lambda$ with $\lambda^2 \leq M^2$. Since the bare charge $Q_{0\text{\textit{st}}}$ can be expressed in terms of $\mathbb{P}_M$ as $ Q_{0\text{\textit{st}}}=(-2)^{-d/2} \Tr\{\Gamma_5\mathbb{P}_M\}$, the staggered spectral expression for the topological susceptibility can now be written:
\beq
\label{stag_sp_topo_susc}
\chi_{\text{\textit{SP}}} = \left(\frac{Z_P^{(\text{\textit{s}})}}{Z_S^{(\text{\textit{s}})}}\right)^2 \frac{\langle Q_{0\text{\textit{st}}}^2\rangle}{V}
= \frac{1}{2^d} 
\frac{\braket{\Tr\{\mathbb{P}_M\}}}{\braket{\Tr{\{\Gamma_5 \mathbb{P}_M\Gamma_5 \mathbb{P}_M\}}}} \frac{\braket{ \Tr{\{\Gamma_5 \mathbb{P}_M\}^2 } }}{V}.
\eeq
Similar expressions can be obtained for higher-order cumulants of the topological charge distribution $P(Q)$, which we parametrize as
\beq\label{cont_expression_b_2n}
b_{2n} \equiv (-1)^n\frac{2}{(2n+2)!} \frac{\braket{Q^{2n+2}}_c}{\braket{Q^2}},
\eeq
where $\braket{Q^k}_c$ denotes the $k^{\text{th}}$-order cumulant of $P(Q)$. Indeed, following the same line of reasoning adopted for the susceptibility, one obtains:
\beq\label{general_expression_stag_sp_b_2n}
b_{2n}^{\text{\textit{SP}}} &=&
(-1)^n\frac{2}{(2n+2)!} \left(\frac{Z_P^{(\text{\textit{s}})}}{Z_S^{(\text{\textit{s}})}}\right)^{2n} \frac{\braket{Q^{2n+2}_{0\text{\textit{st}}}}_c}{\braket{Q^2_{0\text{\textit{st}}}}} \nonumber \\
&=& \frac{(-1)^n}{2^{dn}}\frac{2}{(2n+2)!} \left(\frac{\braket{\Tr\{\mathbb{P}_M\}}}{\braket{\Tr{\{\Gamma_5 \mathbb{P}_M \Gamma_5 \mathbb{P}_M \}}}}\right)^{n} \frac{\braket{ \Tr{\{\Gamma_5 \mathbb{P}_M\}^{2n+2} } }_c }{\braket{ \Tr{\{\Gamma_5 \mathbb{P}_M\}^2 } }}.
\label{b_2n_stag_spectr_proj}
\eeq

\subsection{Choice of the cut-off mass $M$}\label{how_to_set_cut_off}

The choice of the cut-off mass is irrelevant in the continuum limit since, for the index theorem, only zero-modes contribute to the topological charge. However, corrections to the continuum limit depend on $M$ and a prescription to hold the renormalized cut-off $M_R=M/Z_S^{(\text{\textit{s}})}$ fixed as the lattice spacing $a\to0$ is needed to guarantee $O(a^2)$ lattice artifacts~\cite{smit_and_vink,luscher_2}.

Our prescription is to keep the mode density $\braket{\nu}/V$ fixed in physical units for each lattice spacing. Indeed, using leading order chiral perturbation theory and the Banks-Casher relation~\cite{luscher_2} one has:
\beq\label{eq_for_cut_off_M}
\frac{\braket{\nu(M)}}{V} = \frac{2}{\pi} \Sigma M
= \frac{2}{\pi} \Sigma_R M_R, 
\quad \Sigma = -\braket{\bar{\psi}\psi}=-\braket{S_0},
\eeq
where $\nu(M)$ is the number of eigenmodes below $M$ and $\Sigma$ is minus the chiral condensate in the thermodynamic and chiral limit. Since $Z_S^{(\text{\textit{s}})}$ is the renormalization constant for both the chiral condensate and the inverse mass, the product $\Sigma M$ is a renormalization-group-invariant quantity. Therefore, to keep $M_R$ constant, it is sufficient to tune $M$ as a function of $a$ in order to hold $\braket{\nu}/V$ fixed for all lattice spacings.

\section{Numerical tests in the quenched theory}
\label{results}

\subsection{Spectral measure of $\chi$ at $T=0$}\label{subsec_cont_limit_chi}

Simulations were performed using the standard plaquette action on hyper-cubic lattices (with lattice sizes in the range 1.2 - 1.8~fm) for 4 different values of the coupling: $\beta = \{ 5.9, \, 6.0, \, 6.125 , \, 6.25 \}$. For each $\beta$ we collected 300 decorrelated configurations and measured $\chi$ both by the spectral method and with a gluonic definition. The one adopted here is the clover discretization measured after cooling and rounded to the nearest integer, following the lines of Refs.~\cite{DelDebbio:2002xa,Bonati:2015sqt}.

\begin{figure}[!htb]
\centering
\includegraphics[width=0.495\columnwidth, clip]{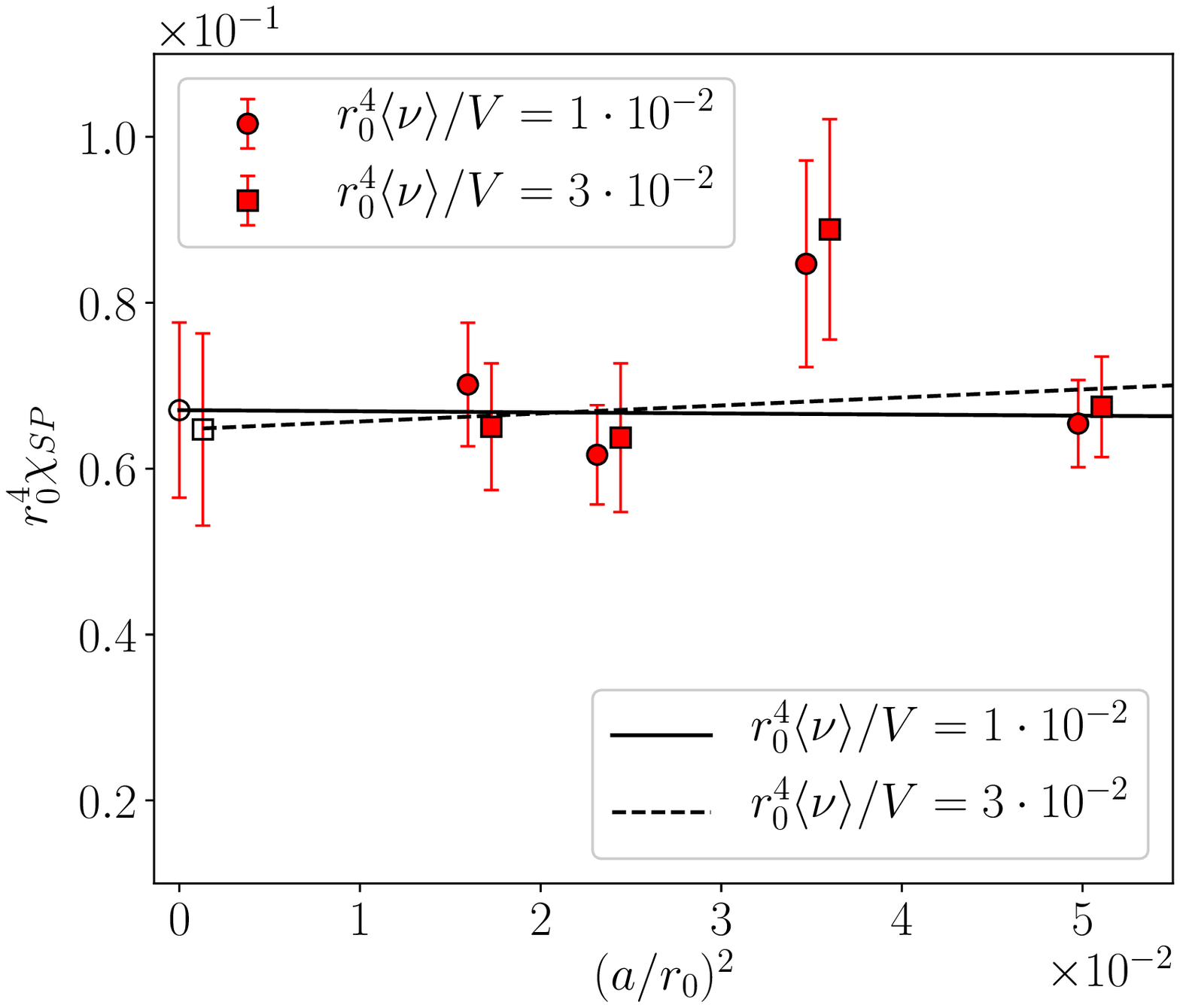}
\includegraphics[width=0.495\columnwidth, clip]{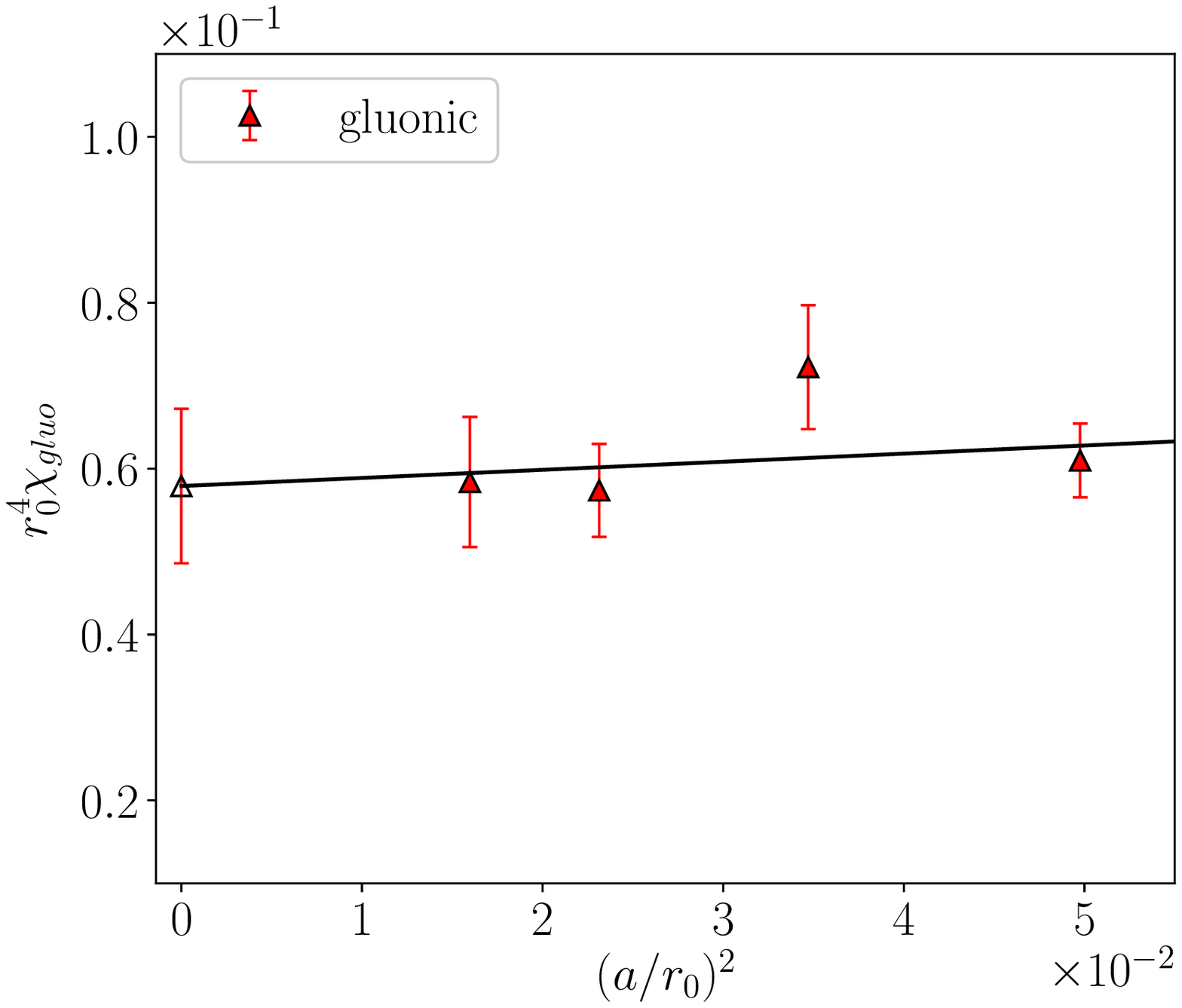}
\caption{Continuum extrapolations of $\chi_{\text{\textit{SP}}}$ and $\chi_{\text{\textit{gluo}}}$ at $T=0$.}
\label{cont_limit_chi_plot}
\end{figure}

In order to extrapolate $\chi_{\text{\textit{SP}}}$ towards the continuum, we considered determinations at two fixed values of the renormalized mass $M_R$, i.e.~at two different values of $\braket{\nu}/V$. As shown in Tab.~\ref{tab_chi_comparison}, the continuum value of $\chi$ obtained with spectral projectors is independent of the choice of $M_R$ and compatible with the gluonic measure within the errors. We also report other determinations of $\chi$ obtained by different fermionic methods: they all agree, within errors, with the staggered spectral results. Continuum extrapolations of $\chi_{\text{\textit{SP}}}$ and $\chi_{\text{\textit{gluo}}}$ are shown in Fig.~\ref{cont_limit_chi_plot}. Lattice artifacts have a weak dependence on the cut-off $M_R$ and their magnitude is comparable to the one of those affecting the gluonic measure (similarly to what happens with Wilson spectral projectors~\cite{luscher_3,Cichy_wilson_spectral_quenched}). 
\begin{table}[!htb]
\begin{center}
\begin{tabular}{ | c || c |}
\hline
& \\[-1em]
Method & $r_0^4 \chi$ \\
\hline
& \\[-1em]
Staggered Spectral Projectors, $M_R=M_1$    & 0.067(11) \\
Staggered Spectral Projectors, $M_R=M_2$    & 0.065(12) \\
Gluonic Definition $+$ Cooling              & 0.058(9)  \\
Wilson Spectral Projectors~\cite{luscher_3} & 0.067(3)  \\
Overlap Operator~\cite{DelDebbio:2004ns}    & 0.059(3)  \\
\hline
\end{tabular}
\end{center}
\caption{Comparison between continuum determinations of $\chi$ for the pure-gauge $SU(3)$ theory. The renormalized cut-offs $M_1$ and $M_2$ correspond respectively to $r_0^4 \braket{\nu}/V = 1 \cdot 10^{-3}$ and $3 \cdot 10^{-3}$, where $r_0\simeq0.5\text{ fm}$ is the Sommer parameter.}
\label{tab_chi_comparison}
\end{table}

\subsection{Spectral measure of $b_2$ at finite $T$}

The fourth cumulant of $P(Q)$ is parametrized through the $b_2$ coefficient, cf.~Eq.~(\ref{cont_expression_b_2n}),
\beq
b_2 = -\frac{1}{12}\frac{\braket{Q^4}-3\braket{Q^2}^2}{\braket{Q^2}}.
\eeq
From Eq.~(\ref{b_2n_stag_spectr_proj}) we get its staggered spectral expression:
\beq
\label{b2_sppr}
b_2^{\text{\textit{SP}}} = -\frac{1}{2^d}\frac{1}{12} \frac{\braket{\Tr\{\mathbb{P}_M\}}}{\braket{\Tr{\{\Gamma_5 \mathbb{P}_M \Gamma_5 \mathbb{P}_M \}}}} \frac{\braket{\Tr{\{\Gamma_5 \mathbb{P}_M\}^{4}}} -3\braket{\Tr{\{\Gamma_5 \mathbb{P}_M\}^{2}}}^2}{\braket{ \Tr{\{\Gamma_5 \mathbb{P}_M\}^2 } }}.
\eeq
The measure of $b_2$ at zero temperature requires quite large statistics since $b_2$ encodes deviations of $P(Q)$ from a Gaussian, which turn out to be quite small~\cite{DelDebbio:2002xa, Bonati:2015sqt,DElia:2003zne, Giusti:2007tu, Ce:2015qha, Bonati:2016tvi}. Thus, we tested the numerical determination of $b_2$ via spectral projectors in the deconfined phase of the $SU(3)$ pure-gauge theory. Indeed, in the high-$T$ regime its value is larger than the $T = 0$ one because it approaches the prediction $b_2 = -1/12$ from the Dilute Instanton Gas Approximation (DIGA).

\begin{figure}[!htb]
\centering
\includegraphics[width=0.495\columnwidth, clip]{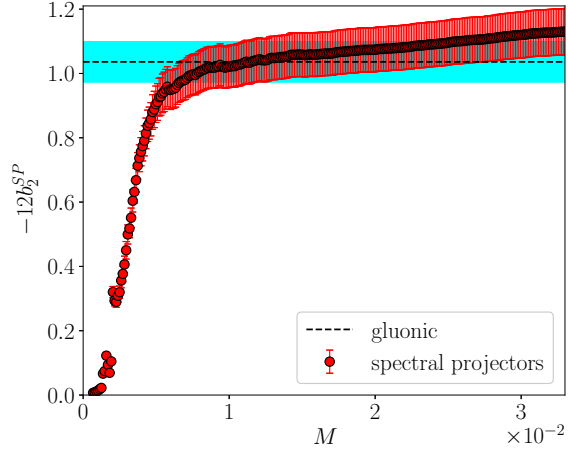}
\caption{Behavior of $- 12 b_2^{\text{\textit{SP}}}$ at $T\simeq338$ MeV with the bare cut-off mass $a M$, compared to the gluonic measure obtained on the same sample.}
\label{b_2_vs_M_fig_beta_6.305}
\end{figure}

In this case we considered a $30^3 \times 10$ lattice with $\beta = 6.305$, corresponding to $T \simeq 338$~MeV~$\simeq 1.145~T_c$.
In Fig.~\ref{b_2_vs_M_fig_beta_6.305} we show results obtained for $b_2^{\text{\textit{SP}}}$ as a function of the bare cut-off mass $M$. \\
In this case we do not perform any continuum limit, thus we do not fix any value of $M$. As it can be appreciated, spectral results are well compatible, over a wide range of $M$, with the gluonic determination of $b_2$ obtained from the same configuration sample.

\section{Conclusions}
\label{conclusions}

In this paper we reported on the main results of \cite{Bonanno:2019xhg}, where the spectral projectors method is extended to staggered fermions. In particular, we derived spectral expressions for the topological susceptibility $\chi$ and for the $b_{2n}$ coefficients, taking into account the renormalization of the fermionic charge and the fourfold degeneracy of the staggered operator, and we tested them in the pure $SU(3)$ gauge theory, both at zero and finite temperature. Our results are in agreement with the ones obtained by other definitions, both gluonic and fermionic. 

In this case lattice artifacts affecting spectral measures are comparable to those affecting gluonic ones, unlike what happens in QCD at physical quark masses. In that case, since non-zero-charge configurations are not efficiently suppressed in the path integral because of the presence of large would-be-zero-modes, topological observables suffer for larger discretization effects compared to the quenched case. Results obtained in~\cite{athenodorou_wilson_spectral_fullQCD}, instead, show that a spectral fermionic definition of $\chi$, matching the same discretization adopted for the Monte Carlo evolution, can strongly reduce such discretization errors and greatly improve the accuracy of the continuum extrapolation.

For this reason, in the near future we plan to apply spectral methods to improve the study of topological observables in high-$T$ QCD with staggered fermions. This investigation will be presented in an upcoming work.

\providecommand{\href}[2]{#2}\begingroup\raggedright\endgroup

\end{document}